\documentclass[twocolumn]{article}
\usepackage{spconf,amsmath,graphicx}
\usepackage{adjustbox}
\usepackage{cite}
 \usepackage{booktabs}
 \usepackage{multirow}

\title{DQ-Whisper: Joint Distillation and Quantization for Efficient Multilingual Speech Recognition}
\name{Hang Shao\sthanks{Equal Contribution.}, Bei Liu$^*$, Wei Wang, Xun Gong, Yanmin Qian\sthanks{Corresponding Author.}}
\address{
  Auditory Cognition and Computational Acoustics Lab\\
  MoE Key Lab of Artificial Intelligence, AI Institute\\
  Department of Computer Science and Engineering, Shanghai Jiao Tong University, Shanghai, China\\
}
\begin{document}
%
\maketitle
\begin{abstract}
    As a popular multilingual and multitask pre-trained speech model, Whisper has the problem of \textit{curse of multilinguality}. To enhance multilingual capabilities in small Whisper models, we propose DQ-Whisper, a novel joint distillation and quantization framework to compress Whisper for efficient inference. Firstly, we propose a novel dynamic matching distillation strategy. Then, a quantization-aware distillation framework is introduced to integrate quantization with distillation. Experimental results on various multilingual datasets show that our suggested distillation approach can effectively enhance the multilingual capabilities of small Whisper models without increasing computational costs. Up to 5.18x reduction in model size is achieved with marginal performance degradation. In addition, quantization is compatible with distillation, which can result in a higher compression rate.
\end{abstract}

\section{Introduction}
Pre-training models on large amounts of data and then fine-tuning them on specific downstream tasks has emerged as a powerful approach for automatic speech recognition (ASR). Several pre-trained audio encoders, such as wav2vec~\cite{schneider2019wav2vec}, wav2vec2~\cite{baevski2020wav2vec}, HuBERT~\cite{hsu2021hubert}, and WavLM~\cite{wavlm} have achieved impressive results in various ASR tasks. However, these models are typically trained in an unsupervised or self-supervised manner, which requires a phase of fine-tuning to enable them to perform well in unseen domains. In contrast, being an encoder-decoder structure, Whisper~\cite{radford2022robust} is pre-trained on a vast corpus of multilingual, weakly supervised data, exhibiting promising robustness and versatility across diverse cross-lingual tasks. 

Recent studies have uncovered the issue of \textit{curse of multilinguality} in Whisper~\cite{multilingualdistilwhisper}, where a significant gap in ASR performance is observed between its larger and smaller variants across a wide range of languages, particularly those with low- and mid-resources. Despite the remarkable ability of the larger Whisper models to cover numerous languages, significant computational and storage requirements make it challenging to deploy them onto devices with restricted resources, such as mobile phones~\cite{hannun2021history}. Enhancing the capabilities of multilingual ASR in smaller Whisper variants is a crucial and urgent task. 

Model compression is a widely utilized technique to reduce model size and speed up inference without sacrificing performance. Previous studies have investigated various methods for self-supervised speech pre-trained models, including knowledge distillation~\cite{chang2022distilhubert,jiao2019tinybert,romero2014fitnets,gong22_interspeech} and network pruning~\cite{wang2022lighthubert,xia2022structured,liu2018rethinking,frankle2018lottery}. For example, DistilHuBERT~\cite{chang2022distilhubert} and FitHuBERT~\cite{fithubert} obtain lite versions of HuBERT by distilling knowledge from a larger teacher model to a more compact student one. On the other hand, DPHuBERT~\cite{dphubert} utilizes structured pruning method to remove unimportant weights and shrink the model size. ~\cite{wang23da_interspeech} introduces a fine-grained pruning strategy for WavLM to achieve a higher compression rate. Similarly, ~\cite{peng_icassp} applies heterogeneous joint pruning using L0 regularization to the wav2vec2 model. These approaches mainly focus on encoder-only architecture, which are not suitable for Whisper. ~\cite{distilwhisper} compresses Whisper by distilling it into a smaller variant using a large-scale pseudo-labelled dataset. ~\cite{multilingualdistilwhisper} proposes a language-specific module to enhance the performance. However, ~\cite{distilwhisper} involves massive training time and ~\cite{multilingualdistilwhisper} incurs extra computational expenses. How to efficiently preserve multilingual capabilities in small Whisper models still remains a big challenge.

In this paper, we propose DQ-Whisper, a novel joint distillation and quantization framework to compress the Whisper model for efficient inference while maintaining multilingual capabilities. Firstly, a novel dynamic matching distillation strategy is proposed. Different from prior works, our method is capable of selecting appropriate layers from teacher network to guide the learning process of student network. Subsequently, we introduce a quantization-aware distillation framework to integrate quantization with distillation. Experimental results on various multilingual datasets show that our suggested distillation approach can effectively enhance the multilingual capabilities of small Whisper models without increasing computational costs. A reduction in model size by up to 5.18x is achieved with only a slight drop in performance. Moreover, integrating distillation with quantization can result in greater compression rate while maintaining the performance. Finally, the evaluation on \textsc{CommonVoice} illustrates that our approach can efficiently condense the Whisper model while preserving its capability for multilingual speech recognition.


\section{Methods}
\subsection{Dynamic Matching Distillation}
Prior works~\cite{distilwhisper, multilingualdistilwhisper} have only used the logit output from prediction layer as the loss function for transferring knowledge from the teacher to the student model. In addition, DPHuBERT~\cite{dphubert} proposes a static layer-wise distillation approach, leveraging the hidden outputs from the 4th, 8th, and 12th HuBERT layers to guide the learning of the student model. However, the fixed layer selection is ad-hoc and heuristic, which might result in sub-optimal outcomes. In this section, we introduce a novel dynamic matching distillation strategy for Whisper that enhances the learning process by merging guidance from both prediction outputs and hidden layers.

\begin{figure}[!t]
  \centering
  \includegraphics[width=.72\linewidth]{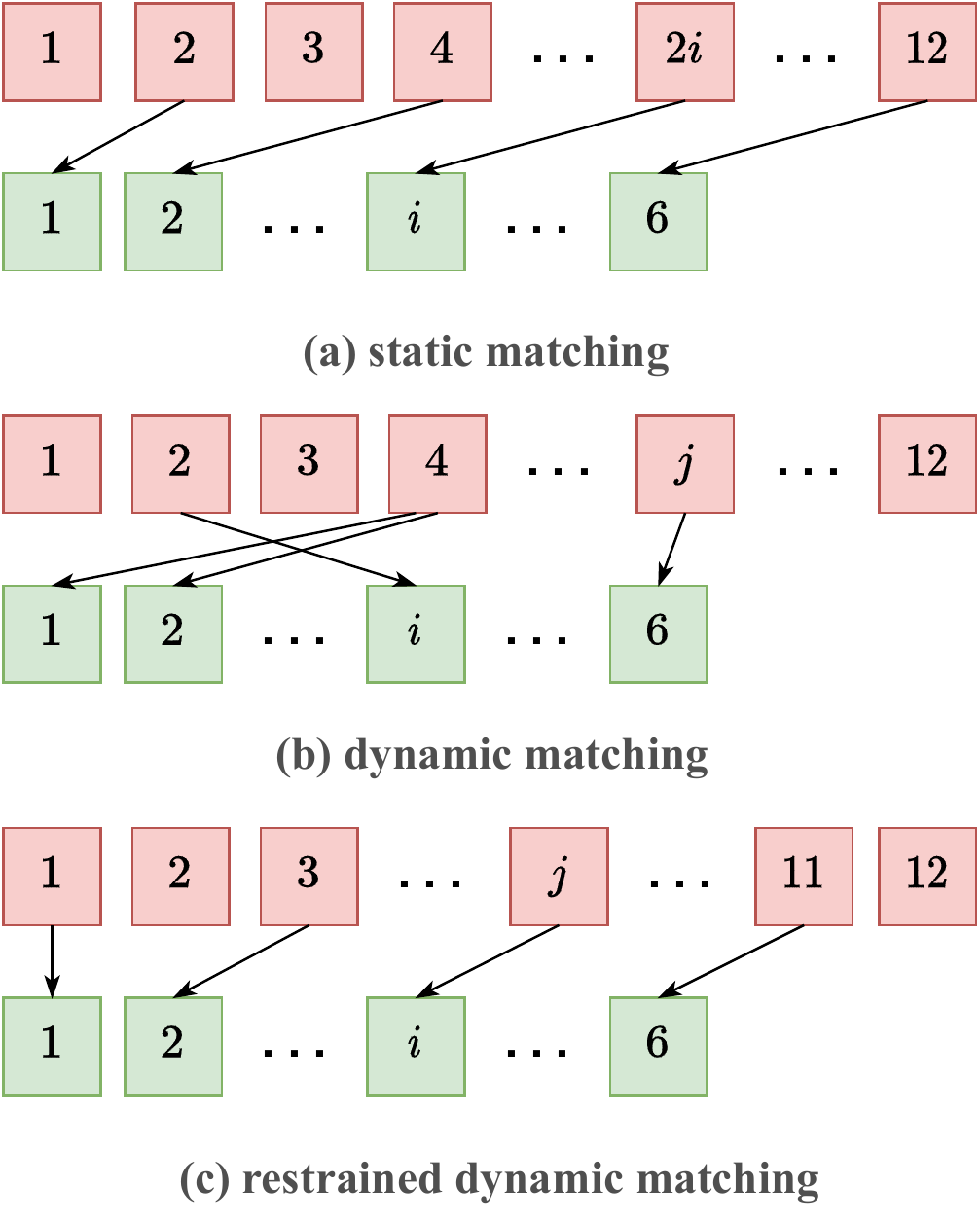}
  \caption{Comparison between three different layer selection methods in hidden layer distillation. {Red} denotes teacher layer and {Green} denotes the student layer.}
  \label{fig:matching_methods}
\end{figure}

\noindent\textbf{Prediction Layer Distillation: } The distillation of the prediction layer is mainly a regularization of the output of the student model and the teacher model, we use the prediction layer of the teacher model to fit the prediction layer of the student model so as to minimize the distribution distance between them, where the loss of the distillation of the prediction layer is:
\begin{equation}
\label{eq:logits_loss}
    L_{pred} = KLD(z^T/t, z^S),
\end{equation}
where $z^T$ and $z^S$ are the logits vectors predicted by the student and teacher respectively and $t$ means the temperature value.

\noindent\textbf{Hidden Layer Distillation: }Whisper consists of multiple Transformer layers, the first step involves matching the student model's layers to those of the teacher model before conducting distillation on the intermediate hidden layers. Different from static layer selection, we develop a dynamic matching strategy to select the appropriate layers across the teacher and student networks. Suppose the student network has $M$ transformer layers and the teacher network has $N$ transformer layers, where $N \ge M$. We need to select $M$ layers from the $N$ layers in the teacher network to correspond with the student network. Then a function $N_j=f(M_i)$ is defined to match the layer $M_i$ of the student network with the layer $N_j$ of the teacher network, that is, the layer $M_i$ of the student network learns the knowledge of the layer $N_j$ of the teacher network. Formally, the student can acquire knowledge from the teacher by minimizing the following objective:
\begin{equation}
\label{eq:hidden_loss}
    L_{hidn} = \frac{1}{M}\sum_{i=1}^{M}\lambda_{i}MSE(H_i^S,H_{f(i)}^T),
\end{equation}
where $H^S_i$ and $H^T_{f(i)}$ denote the hidden layer feature vectors of layer $i$ and layer $f(i)$ of the student model and teacher model respectively, and $\lambda_{i}$ is the hyper-parameter that represents the importance of the $i$-th layer's distillation. Specifically, two different dynamic mapping function $f(i)$ are designed as follows:
\begin{itemize}
    \item \textbf{Dynamic Matching (DM)}: We select the layer $f(i)$ with the minimum mean squared error (MSE) from the teacher model for each layer $i$ of the student model.
    \begin{equation}
    \label{eq:dynamic_no_limit}
        f(i)=\arg\min_{j}||W_AH^T_j-H^S_i||^2,
    \end{equation}
    where $W_A$ is a linear transformation matrix.
    \item \textbf{Restrained Dynamic Matching (RDM)}: Based on eq.~\eqref{eq:dynamic_no_limit}, we propose an improved version of DM by restricting $f(i)$ to be a monotonically increasing function. Consequently, $f(i)$ is always rigorously $\geq i$.
\end{itemize}

\subsection{Joint Distillation and Quantization}
Recent studies have shown that different compression methods can complement each other. For example, DPHuBERT~\cite{dphubert} indicates that integrating knowledge distillation with network pruning leads to improved outcomes and a higher compression rate. Similarly, ~\cite{mixedgpt} discovers that pruning can be effectively combined with network quantization in large language models (LLMs). Inspired by these findings, we devise a novel framework that combines distillation and quantization to push the compression limits.

\noindent\textbf{Uniform Quantization: }For n-bit uniform quantization, the set of integers N that can be selected by the quantization matrix in the model is defined as:
\begin{equation}
  \boldsymbol{N} \in\left\{0, \pm 1, \pm 2, \ldots, \pm 2^{n-1}\right\},
  \label{eq:uniform_table}
\end{equation}
For Transformer, we can construct a quantization table for each of its linear layers. For example, for the $l$-th linear layer of the model, the value range of the quantized parameter $Q^l$ is shown as follows:
\begin{equation}
  \boldsymbol{Q}^{l}=\alpha^{l} \boldsymbol{N}^{l} \in\left\{0, \pm \alpha^{l}, \ldots, \pm \alpha^{l}\left(2^{n_l-1}\right)\right\},
  \label{eq:uniform}
\end{equation}
where $\alpha^{l}$ represents the scaling factor of $l$-th layer, it flexibly scales the parameter coverage of the current layer after quantization according to the original parameters, $n_{l}$ is the quantization precision number of $l$-th layer. In the following experiments, we set n to 8.

\noindent\textbf{Quantization-Aware Distillation: }Following the approach described in Section 2.1, we design a novel quantization-aware distillation framework where quantization loss is adopted to guide layer selection during dynamic matching. As eq.~\eqref{eq:dynamic_no_limit} illustrates, the initial matching criterion aims to minimize. By integrating quantization, we suggest that quantization loss could serve as a guide for choosing which layers to distill, identifying those that are most suitable for quantization from the teacher network. 

Firstly, uniform quantization is applied to student network. The quantization loss $L_{quan}$ of layer $l$ of the student network is defined as follows:
\begin{equation}
\label{eq:quan_loss}
    L_{quan}^l = |W^l_S-Q^l_S| \approx |W_1W_TW_2-Q^l_S|,
\end{equation}
where $W_1$ and $W_2$ are linear transformation matrices. $W_S$ and $Q_S$ are the weight of the student network before and after quantization, respectively. $l$ denotes the $l$-th layer.

Then, we guide the selection of distillation layers by the quantization loss $L_{quan}$. For each layer $i$ in the student network, the layer $f(i)$ is chosen by minimizing the quantization loss derived from the teacher network. The mapping function $f(i)$ is presented below:
\begin{equation}
\label{eq:map_function}
    f(i)=\arg \min_j{|W_1W_T^jW_2-Q^i_S|},
\end{equation}

\begin{table*}[!t]
\centering
\caption{
CER (\%), Model Size and Compression ratio comparison of different setups on in-house Chinese datasets.
\texttt{Logits} denotes distillation of only logits layer, and \texttt{uniform} indicates uniform quantization of all layers. The two numbers in this column of IHC1/IHC2 are the CERs (\%) of IHC1 and IHC2 test sets respectively. \texttt{Avg} indicates the averaged CER (\%) on these two test sets.
\texttt{Compres.} is the compression ratio compared to the teacher model.
\texttt{Size} is the storage of the model on the disk. \texttt{KD} means knowledge distillation. \texttt{DQ} is joint distillation and quantization.
Small$\rightarrow$Base/Tiny means that Small is compressed to Base or Tiny size.
}
\begin{tabular}{l|cccc|cccc}
\toprule
\multirow{2}{*}{Method} & \multicolumn{4}{c|}{Small $\rightarrow$ Base} & \multicolumn{4}{c}{Small $\rightarrow$ Tiny} \\
& IHC1/IHC2 & Avg & Size & Compres. & IHC1/IHC2 & Avg & Size & Compres.  \\
\midrule
Teacher & 21.6/33.1 & 27.4 & 461MB & 1.0$\times$ & 21.6/33.1 & 27.4 & 461MB & 1.0$\times$ \\
\midrule
Student & 29.1/39.1 & 34.1 & \multirow{4}{*}{139MB} & \multirow{4}{*}{3.32$\times$} & 35.1/42.2 & 38.7 & \multirow{4}{*}{72MB} & \multirow{4}{*}{6.40$\times$} \\
+ KD logits & 25.0/37.9 & 31.5 &  &  & 30.0/39.3 & 34.7 &  &  \\
++ dynamic matching & 24.5/38.9 & 31.7 & &  & 30.4/38.9 & 34.7 &  &  \\
++ restrained dynamic matching & \textbf{23.2}/\textbf{36.5} & \textbf{29.9} & &  & \textbf{29.2}/\textbf{38.0} & \textbf{33.6} & & \\
\midrule
+ DQ logits uniform & 25.9/38.3 & 32.1 & \multirow{3}{*}{89MB} & \multirow{3}{*}{\textbf{5.18$\times$}} & 30.5/39.6 & 35.1 & \multirow{3}{*}{44MB} & \multirow{3}{*}{\textbf{10.48$\times$}} \\
++ dynamic matching & 25.1/38.8 & 32.0 &  &  & 30.6/39.9 & 35.3 &  & \\
++ restrained dynamic matching & \textbf{23.9}/\textbf{36.6} & \textbf{30.3} &  &  & \textbf{28.2}/\textbf{38.3} & \textbf{33.3} &  &  \\
\bottomrule
\end{tabular}
\label{tab:Compression}
\end{table*}

Finally, the distillation and quantization are jointly trained together via the following loss function:
\begin{equation}
\label{eq:Kdloss}
    L_{kd} = L_{pred} + L_{hidn},
\end{equation}
\begin{equation}
\label{eq:total_loss}
    L_{model} = \alpha L_{kd} + \gamma L_{quan} + (1 - \alpha)CE(\hat{y_s},y),
\end{equation}
where $L_{kd}$ represents prediction and hidden layer distillation loss. $L_{quan}$ denotes the quantization loss. $CE$ means cross-entropy loss. $\hat{y_s}$ and $y$ are the output probability distribution of the student model and the label, respectively.

In contrast to prior approaches, our quantization-aware distillation method for Whisper integrates quantization with dynamic matching distillation, achieving not only a higher compression rate but also preserving the multilingual abilities of the student models.

\section{Experiment}

\subsection{Experimental Setup}
\noindent\textbf{Datasets: }Our training dataset is from the 450 hours in-house Chinese dataset, and the test set is from a 10 hours in-house Chinese test set IHC1 and 15 hours IHC2. For multilingual datasets, we use \textsc{Commonvoice} \cite{ardila2019common} and selected four languages for our experiment: German~(de), Spanish~(es), Catalan~(ca) and French~(fr), totaling 1393 hours. The details are shown in Table~\ref{tab:Commonvoice}.

\begin{table}[ht]
\centering
\caption{
Details of four languages used in the experiments from the \textsc{Commonvoice}}
\begin{tabular}{l|c|c|c}
\toprule
Language & \#Spk. & \#Utt. & Duration  \\
\midrule
German~(de) & 9,191 & 227,086 & 364 hr\\
Spanish~(es) & 9,045 & 168,598 & 254 hr\\
Catalan~(ca) & 4,667 & 230,993 & 361 hr\\
French~(fr) & 10,035 & 286,105 & 414 hr\\
\bottomrule
\end{tabular}
\label{tab:Commonvoice}
\end{table}

\noindent\textbf{Model: }Due to the limited GPU memory, we adopt \textit{whisper-small} as the teacher model in the experiments, which consists of 12 encoder layers and 12 decoder layers with 3072 hidden units. Each encoder and decoder layer is a transformer block with 12 heads of 768 dimension self-attention~\cite{vaswani2017attention}. Meanwhile, \textit{whisper-base} and \textit{whisper-tiny} are employed as the student models.

\subsection{Implementation Details}
We use an 80-channel log-magnitude Mel spectrogram with 25ms window length computed every 10ms as inputs of audio encoder. Plus, SpecAugment~\cite{park2019specaugment} as a
data augmentation policy is used during model training. The Adam~\cite{kingma2014adam} optimizer is adopted with 3e-5 initial learning rate and 10,000 warm-up steps. In eq.~\eqref{eq:total_loss} $\alpha$ is 0.5 and $\gamma$ is 1.0. In eq.~\eqref{eq:hidden_loss} $\lambda_i$ is 1.0 for all $1\le i \le M$. We use 8-bit quantization. Owing to the constraints of GPU memory, we freeze the parameters of encoder in the training process and trained only the decoder part. We use the same byte-level BPE text tokenizer used in GPT-2~\cite{radford2019language,sennrich2015neural} 
for the multilingual models. All models are trained with in-house speech recognition toolkit Pytorch-asr and ESPnet~\cite{watanabe2018espnet} until convergence.

\section{Results and Analysis}
\subsection{Evaluation on Dynamic Matching Distillation}
Due to the limited GPU memory, we adopt \textit{whisper-small} as the teacher model, with \textit{whisper-base} and \textit{whisper-tiny} serving as the student models. Firstly, both teacher and student models are fine-tuned on 450h in-house Chinese dataset.

From Table~\ref{tab:Compression}, it can be clearly observed that \textit{whisper-base} and \textit{whisper-tiny} still exhibit poor performance in out-of-domain tasks, with a gap in CER of up to 11\% compared to \textit{whisper-small}. This reflects the problem of \textit{curse of multilinguality} in Whisper again. By applying logit distillation, the performance of \textit{whisper-base} can be reduced from 34.1\% to 31.5\%. When incorporating hidden layer distillation, CER can be further decreased to 29.9\%, nearly matching whisper-small's performance but with a model size that is 3.32x smaller. Similar phenomena can be seen in \textit{whisper-tiny}. For example, 10.3\% relative improvements can be achieved when using prediction layer distillation. The integration of hidden layer distillation can further lower CER to 33.6\%, yielding a 6.4x reduction in parameters.

The above analysis reveals that our suggested distillation approach can effectively enhance the multilingual capabilities of small Whisper models without increasing computational costs. Moreover, hidden layer distillation is compatible with logit distillation, and their integration can further improve results. This is due to the important information spread in both the intermediate and prediction layers of Whisper models. It's also noted that restrained dynamic matching gives the best performance, highlighting the significance of selecting the appropriate layers during hidden layer distillation. In addition, a monotonically increasing order proves to be the most effective for Whisper models.

\begin{table}[!t]
    \centering
    \caption{Performance CER(\%) comparison of different compression strategies. \texttt{quant\_KD} means applying quantization and distillation separately. \texttt{DQ} denotes our proposed joint distillation and quantization method. \texttt{Avg} indicates the averaged CER (\%) on IHC1 and IHC2. \texttt{Size}(MB) indicates the storage of the model on the disk.}
    \begin{adjustbox}{max width=.46\textwidth}
    \begin{tabular}{l|c|c|c|c}
        \toprule
        System & \multicolumn{1}{c|}{IHC1/IHC2} & \multicolumn{1}{c|}{Avg} & \multicolumn{1}{c|}{Size} & \multicolumn{1}{c}{Compress.} \\
        \midrule
        Small & 21.6/33.1 & 27.4 & 461MB & 1.0$\times$ \\
        \midrule
        Base & 29.1/39.1 & 34.1 & 139MB & 3.32$\times$ \\
        + quant\_KD & 27.3/39.5 & 33.4 & 89MB & 5.18$\times$ \\
        + DQ & \textbf{23.9/36.6} & \textbf{30.3} & \textbf{89MB} & \textbf{5.18$\times$} \\
        \midrule
        Tiny & 35.1/42.2 & 38.7 & 72MB & 6.40$\times$ \\
        + quant\_KD & 34.8/39.9 & 37.3 & 44MB & 10.48$\times$ \\
        + DQ & \textbf{28.2/38.3} & \textbf{33.3} & \textbf{44MB} & \textbf{10.48$\times$} \\
        \bottomrule
    \end{tabular}
    \end{adjustbox}
    \label{tab:quant_kd}
\end{table}

\subsection{Evaluation on Joint Distillation and Quantization}
In this section, we examine the effect of joint distillation and quantization scheme. In the experiments, all layers of the student model in the decoder are quantized to 8-bit integers. As Table~\ref{tab:Compression} displays, we can see that the combination of  distillation and quantization can achieve a much higher compression rate. For instance, the sizes of \textit{whisper-base} and \textit{whisper-tiny} models decrease from 139MB/72MB to 89MB/44MB, achieving compression ratios of 5.18x and 10.48x. On the other hand, the performance is nearly identical to the best distilled model. This indicates that distillation is compatible with quantization. The combination of them can not only yield a promising compression rate, but also maintain the performance. In addition, layer selection based on quantization loss exhibits the best results, demonstrating the effectiveness and superiority of our proposed method.

\subsection{Ablation Study}
In this section, we analyze the effects of joint distillation and quantization approach on performance. Table~\ref{tab:quant_kd} clearly demonstrates that our joint training strategy significantly outperforms the two-stage compression method. Compared to separate quantization and distillation, our DQ can achieve 9.2\%/10.7\% CER relative improvements for \textit{whisper-base} and \textit{whisper-tiny} respectively, while maintaining the same compression rate. This indicates that the joint training of distillation and quantization leads to more effective knowledge transfer and reduces the performance gap between teacher and student networks. Additionally, separate compression involves a two-step training process, which requires more computational resources and time, highlighting the efficiency of our approach.

\begin{table}[!t]
    \centering
    \caption{Performance WER(\%) comparison of different setups on \textsc{Commonvoice} corpus. \texttt{Base\_KD} indicates distillation of the teacher model to the base level. \texttt{Base\_KDQ} denotes the distillation of the teacher model to the base level while quantizing it with 8bit. \texttt{Avg} indicates the averaged WER (\%) on these four test sets. \texttt{Size}(MB) indicates the storage of the model on the disk.}
    \begin{adjustbox}{max width=.46\textwidth}
    \begin{tabular}{l|c|c|c|c|c|c}
        \toprule
        System & \multicolumn{1}{c|}{de} & \multicolumn{1}{c|}{es} &\multicolumn{1}{c|}{ca} & \multicolumn{1}{c|}{fr} & \multicolumn{1}{c|}{Avg} & \multicolumn{1}{c}{Size}\\
        \midrule
        1.Small & 15.6 & 16.8 & 26.2 & 28.8 & 21.8 & \multirow{2}{*}{461}\\
        2.+fine-tuning & 12.0 & 12.6 & 18.1 & 23.1 & 16.5 \\
        3.+8bit quan & 13.2 & 13.0 & 20.8 & 25.4 & 17.9 & 315 \\
        4.+4bit quan & 16.4 & 17.5 & 27.3 & 30.6 & 23.1 & 241 \\
        \midrule
        5.Base & 26.5 & 25.2 & 39.7 & 39.9 & 32.8 & \multirow{3}{*}{139} \\
        6.+fine-tuning & 16.4 & 15.1 & 23.5 & 27.9 & 20.7 &  \\ 
        7.Base\_KD & \textbf{13.9} & \textbf{13.6} & \textbf{21.2} & \textbf{25.6} & \textbf{18.6} & \\
        8.Base\_DQ & 14.0 & 13.9 & 21.3 & 25.7 & 18.7 & \textbf{89} \\
        \midrule
        9.Tiny & 43.3 & 39.5 & 61.6 & 57.9 & 50.6 & \multirow{3}{*}{72} \\
        10.+fine-tuning & 20.7 & 19.2 & 26.7 & 32.1 & 24.7 & \\ 
        11.Tiny\_KD & \textbf{16.5} & \textbf{13.1} & \textbf{23.8} & \textbf{30.4} & \textbf{21.0} &  \\
        12.Tiny\_DQ & 16.7 & 13.3 & 23.9 & 30.6 & 21.1 & \textbf{44}\\
        \bottomrule
    \end{tabular}
    \end{adjustbox}
    \label{tab:multilingual}
\end{table}

\subsection{Evaluation on Multilingual Corpus}
In this section, we evaluate the proposed methods on the multilingual dataset, \textsc{CommonVoice}. Table~\ref{tab:multilingual} presents the experimental results obtained from four multilingual tasks, including German, Spanish, Catalan and French. The first row displays the results of decoding the teacher model, \textit{whisper-small}, directly in different languages. The second row shows the testing results in various languages after fine-tuning the model with the corresponding target languages. We can see that the performance can significantly improved by fine-tuning Whisper for target languages. The third and fourth rows respectively show the results of 8-bit and 4-bit quantization for \textit{whisper-small}. The fifth row lists the results of decoding the student model, \textit{whisper-base}, directly in different languages. The sixth row displays the results of testing the \textit{whisper-base} model after fine-tuning it in the corresponding languages. Lastly, the seventh row demonstrates the results obtained from \textit{whisper-small} distillation to \textit{whisper-base} using the best configuration distillation method mentioned in this paper.

Furthermore, the eighth row shows the results of joint distillation and quantization proposed in this paper. It is noted that there is a negligible degradation in WER performance compared to the fifth row. However, the model size is reduced from 139MB to 89MB, resulting in an increased compression ratio from 3.32x to 5.18x. These findings affirm that our approach can efficiently condense the Whisper model while preserving its capability for multilingual recognition. Similar conclusion can also be drawn from \textit{whisper-tiny} model.

\section{Conclusion}
In this paper, to address the problem of \textit{curse of multilinguality} in Whisper, we introduce DQ-Whisper, a new framework combining distillation and quantization to efficiently compress the Whisper model for effective inference, while preserving its multilingual capabilities. Firstly, we devise a novel dynamic matching distillation approach which enables the selection of specific teacher network layers to guide the student network's learning. Then, Furthermore, we employ a quantization-aware distillation method, seamlessly merging quantization and distillation. Experimental results on various multilingual datasets show that our suggested distillation approach can substantially enhance the multilingual abilities of smaller Whisper models without additional computational burden, attaining a reduction in model size by as much as 5.18x with a negligible decline in performance. Furthermore, the combination of distillation and quantization yields a higher compression ratio while maintaining the performance. Finally, Our evaluation on \textsc{CommonVoice} confirms that DQ-Whisper can effectively compress the Whisper model while preserving its capability for multilingual speech recognition.

\section{Acknowledgement}
This work was supported in part by China NSFC projects under Grants 62122050 and 62071288, in part by Shanghai Municipal Science and Technology Commission Project under Grant 2021SHZDZX0102.

\newpage
\bibliographystyle{IEEEtran}
\bibliography{mybib}

\end{document}